# Quartz as an Accurate High-Field Low-Cost THz Helicity Detector

Maximilian Frenzel[1], Joanna M. Urban[1], Leona Nest[1], Tobias Kampfrath[1,2],
Michael S. Spencer[1], Sebastian F. Maehrlein[1,†]

[1] *Fritz Haber Institute of the Max Planck Society,*
   *Department of Physical Chemistry, 14195 Berlin, Germany*

[2] *Freie Universität Berlin, Department of Physics, 14195 Berlin, Germany*

[†]Corresponding author. Email: maehrlein@fhi-berlin.mpg.de

**Abstract**

The advent of high-field THz sources has opened the field of nonlinear THz physics and unlocked access to fundamental low energy excitations for ultrafast material control. Recent advances towards controlling and employing chiral excitations, or generally angular momentum of light, not only rely on the measurement of undistorted intense THz fields, but also on the precise knowledge about sophisticated THz helicity states. A recently reported and promising detector material is α-quartz. However, its electrooptic response function and contributing nonlinear effects have remained elusive. Here, we establish z-cut α-quartz as a precise electrooptic THz detector for full amplitude, phase and polarization measurement of intense THz fields, all at a fraction of costs of conventional THz detectors. We experimentally determine its complex detector response function, which is in good agreement with our model based on predominantly known literature values. It also explains previously observed thickness-dependent waveforms. These insights allow us to develop a swift and reliable protocol to precisely measure arbitrary THz polarization and helicity states. This two-dimensional electrooptic sampling (2D-EOS) in α-quartz fosters rapid and cost-efficient THz time-domain ellipsometry, and enables the characterization of polarization-tailored fields for driving chiral or other helicity-sensitive quasiparticles and topologies.

**Introduction**

THz sources with peak field strengths in the ~1 MV/cm regime, employing optical rectification in $LiNbO_3$[1], difference frequency generation[2], large-area spintronic emitters[3] and accelerator-based facilities[4], are becoming more widely accessible. This development has enabled the selective drive of low-energy excitations such as phonons[5,6], magnons[7] or other quasiparticles, thereby allowing for ultrafast control over material properties and non-equilibrium material design towards light-induced superconductivity[8], ferroelectricity[9], ferromagnetism[10] and spin-dynamics[7,11]. However, despite large improvements in THz generation, the detection of intense single-cycle THz fields without distortions has remained challenging[12,13].

Field-resolved THz detection provides precise frequency resolution of amplitude and phase of the light field. This feature is crucial for, e.g., THz time-domain spectroscopy (THz-TDS)[14], THz emission spectroscopy[13] and state-of-the-art experiments involving THz high-harmonic generation in topological insulators[15], graphene[16] or superconducting cuprates[17]. Moreover, emerging field-driven effects, e.g., for ultrafast control of topological[18] or chiral[19-21] material properties, are inherently sensitive to the carrier-envelope phase (CEP) and polarization (e.g., helicity) of the driving THz pulse. Full vectorial THz-field characterization is required for the precise detection of arbitrary THz polarization states. This information constitutes the basis for THz time-domain ellipsometry, which allows for the characterization of tensorial dielectric properties in opaque[22], anisotropic materials[23], and transient metamaterials[24], where traditional THz-TDS faces limitations. Another application is THz circular-dichroism spectroscopy, which has been applied in chiral nanostructures and molecular assemblies[25], thermoelectric solids[26], or bio-relevant systems such as DNA[27], and living cancer cells[28]. However, partly due to the difficulty of precise polarization-resolved THz detection, THz time-domain ellipsometry and circular-dichroism spectroscopy have not been widely adopted yet.

The common technique to detect phase-stable THz fields is electrooptic sampling (EOS)[29]. Here, the incident THz pulse induces a change in birefringence proportional to the THz electric field in a nonlinear crystal like ZnTe[30] or GaP[31], which can be stroboscopically sampled by a visible (VIS) or near-infrared (NIR) sampling pulse as a function of time delay $t$. However, the measured instantaneous signal $S(t)$ is, in general, not simply proportional to the instantaneous field $E(t)$ because of noninstantaneous features such as phonon resonances and velocity mismatch of the THz and sampling pulse. Within linear response theory and in frequency space, a response function $h$ connects $S$ and $E$ at THz frequencies $\Omega$ via $S(\Omega) = h(\Omega)E(\Omega)$ and captures the frequency dependence of the nonlinear susceptibility $\chi^{(2)}$, which can be strongly modulated by phonons[29], and non-local effects, such as phase mismatch between THz- and sampling pulse[32,33].

For (110)-oriented zincblende-type electrooptic crystals such as ZnTe (110), resolving the polarization state of THz pulses typically requires rotation of the detector crystal and sampling pulse polarization[34]. Unfortunately, such measurements can be easily polluted by inhomogeneities of the detector crystal, birefringence effects, or inaccurate rotation axes. On the other hand, (111)-oriented zincblende crystals enable polarization state retrieval by simply modulating the sampling pulse polarization by using, e.g., a photoelastic modulator[35] or employing a dual detection scheme based on two balanced detections[36]. Nonetheless, the specific detector requirements and additional experimental effort have limited the application of polarization-resolved EOS so far.

Extending these concepts to highly intense THz fields poses extra challenges, since they can lead to distorted signals in conventional EOS crystals, such as ZnTe or GaP, which include over-rotation[12] or higher-order nonlinearities such as the THz Kerr effect[37,38]. This aspect means that the amplitude and phase of intense THz fields cannot be reliably extracted within the linear response. However, attenuating the THz fields by using, e.g., wiregrid polarizers or filters might induce additional spectral distortions[39].

Here, we focus on z-cut α-quartz, which is a widely used substrate material for THz-TDS due to its high THz transparency[14] and in-plane optical isotropy. It recently attracted attention as a promising nonlinear THz material[40], i.e., as broadband THz emitter via optical rectification[41] or as THz detector via EOS[42]. In fact, its electrooptic coefficient, $r_{11} = 0.1 - 0.3$ pm/V[43], is about an order of magnitude smaller than $r_{41} = 4$ pm/V of ZnTe[30], thereby moving nonlinear EOS responses to much higher THz field amplitudes. Its large bandgap and optical transparency allow for a broad dynamic range and high damage threshold. Moreover, α-quartz is widely available at 2 orders of magnitude lower cost than typical EOS crystals. However, there are significant drawbacks that prevented the reliable use of quartz for THz detection so far. In particular, the response function $h$ has been unknown, and its peculiar thickness dependence lead to the open question regarding bulk versus surface $\chi^{(2)}$ contributions[42]. Likewise, the polarization-sensitivity has remained mostly unexplored.

In this work, we experimentally measure the quartz response function and model it predominantly based on known literature values. We show that arbitrary THz polarization states can be measured by a simple and time-efficient method utilizing only two EOS measurements with different sampling pulse polarizations. The latter is achieved by a simple rotation of a half-waveplate (HWP) in the VIS spectral range. As a textbook example for time-domain ellipsometry, we determine the birefringence of y-cut quartz as commonly used for commercial THz waveplates. We find that the transmitted single-cycle pulses exhibit complex polarization states in the highly polychromatic regime[44], which cannot be described by a single polarization ellipse, Jones vector or set of Stokes parameters. Our study establishes z-cut α-quartz as a reference detector for amplitude, phase, and arbitrary polarization states of THz fields exceeding 100 kV/cm, fostering cost-efficient high-field THz time-domain ellipsometry and tailoring helical THz driving fields for ultrafast material control.

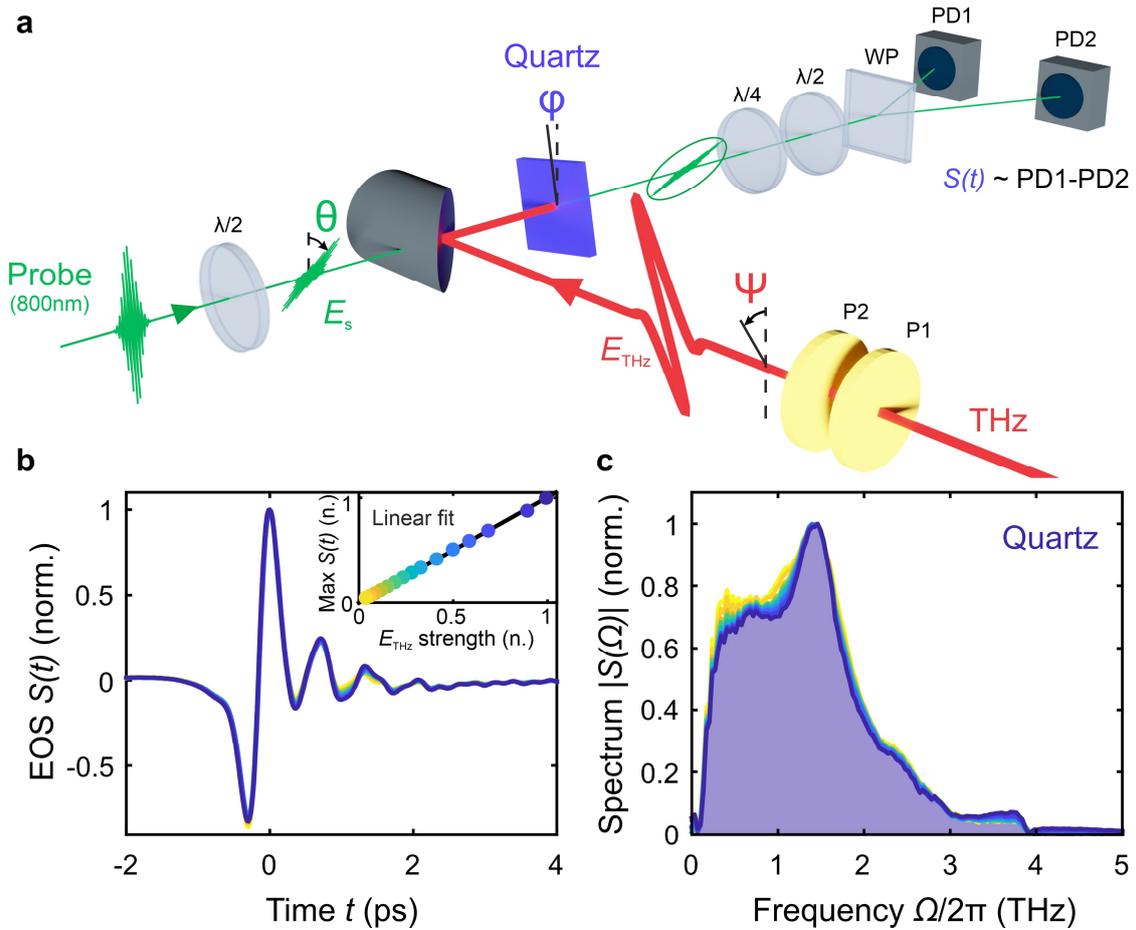

**Fig. 1 | Electrooptic sampling in quartz and its THz fluence dependence. a** Experimental setup: THz pulses are generated via optical rectification (OR) in LiNbO₃. The THz pulse induces a refractive index change in quartz, leading the sampling pulse to acquire ellipticity. This ellipticity is read out as signal $S(t)$ as a function of time delay $t$ in a balanced detection scheme. $S$ is related to the incident THz field $E_{THz}$ via the complex detector response function $h_Q$. **b** EOS in quartz (z-cut, 50 μm thickness) for different THz fluences, normalized to the $t = 0$ peak EOS values. Inset: Linear dependence of peak $S(t)$ on peak $E_{THz}$. **c** $S(\Omega)$ amplitude spectrum via Fourier transform of EOS signals $S(t)$ in (b) normalized to spectral peak amplitude.

## Results

### *Experimental Setup*

Intense single-cycle THz fields (1.3 THz center frequency, 1.5 THz full width at half maximum (FWHM)) with peak fields exceeding 1 MV/cm are generated by tilted-pulse-front optical rectification in LiNbO₃ (Ref. 1). The THz field strengths or its linear polarization angle $\psi$ relative to the vertical direction in the lab frame are altered using a THz polarizer pair, P1 and P2 in **Fig. 1a**. The THz field-induced birefringence in the EOS crystal is probed by synchronized VIS sampling pulses (800 nm center wavelength, ~20 fs duration) using a balanced detection scheme. The sampling pulse's incident linear polarization is set to arbitrary angles $\theta$ by a broadband VIS HWP. We measure EOS in a ZnTe (110) crystal (10 μm thickness) and various z-cut α-quartz plates with thicknesses of 35, 50, 70, and 150 μm as a function of sampling pulse polarization $\theta$, THz polarization $\psi$, and the crystal's azimuthal angle $\phi$ at normal incidence (see **Fig. 1a**). Finally, we also trace the THz field after collimated transmission through highly birefringent y-cut α-quartz (700 μm thickness), which corresponds to a commercial quarter-wave plate (QWP) for 2.2 THz.

*Electrooptic Response Function*

We first confirm the linear response function relation. **Fig. 1b** shows the measured THz-induced birefringence signals $S(t)/S_{\mathrm{max}}$ in 50 μm quartz for different THz peak fields. The induced birefringence scales linearly with the THz electric field strength (see inset of **Fig. 1b**), confirming a linear electrooptic effect as recently observed by Balos *et al.*[42]. The normalized time- and frequency-domain shapes (see **Fig. 1c**) do not change substantially for different THz fluences, ruling out over-rotation effects and demonstrating that the higher-order nonlinearities[40] (e.g. <1.5 THz) are small for THz fields on the order of 1 MV/cm. This finding confirms that quartz can reliably sample THz electric fields $\geq 0.1$ MV/cm within the linear-response regime.

To experimentally extract the linear response function of 50 μm quartz, we compare the quartz EOS signal $S_{\mathrm{Q}}$ with the signal $S_{\mathrm{ZnTe}}$ from 10 μm ZnTe, whose response function $h_{\mathrm{ZnTe}}$ is known[32] (see **Fig 2a**). To avoid nonlinear distortions, the THz power for ZnTe was attenuated by the THz polarizer pair by a factor of ~40. We Fourier transform these traces and extract the quartz response using $h_{\mathrm{Q}} = h_{\mathrm{ZnTe}}(S_{\mathrm{Q}}/S_{\mathrm{ZnTe}})$ in the frequency-domain. The amplitude and phase of $h_{\mathrm{Q}}$ are shown as blue dots in **Figs. 2b** and **2d**, respectively, demonstrating that the quartz response covers the full 0.1-4 THz bandwidth of the LiNbO₃ source without gaps. However, it contains a substantial frequency dependence in the form of modulations with a frequency spacing of ~1.4 THz as well as an enhancement at low frequencies <0.9 THz and at around 3.9 THz.

*Modelling*

To understand the experimental response function $h_{\mathrm{Q,exp}}(\Omega)$ of 50 μm quartz, we model the response $h_{\mathrm{calc}}$ as function of THz frequency $\Omega$ by extending the formalism of Ref. 32 and use:

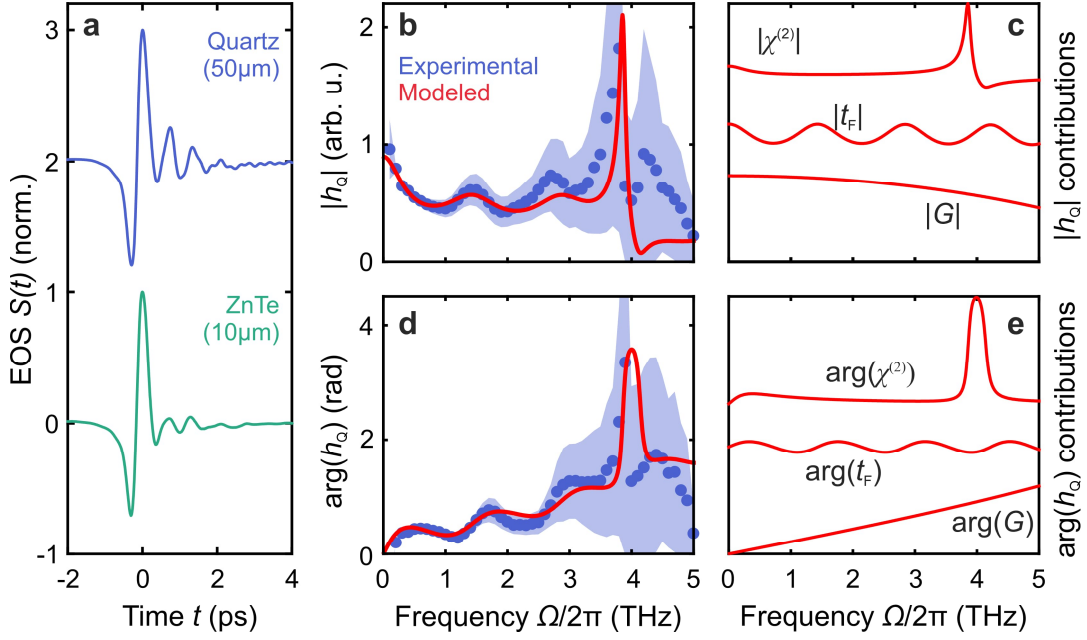

**Fig. 2 | Experimentally measured and calculated detector response. a** Normalized EOS signal $S(t)$ in quartz (50 µm thickness) and ZnTe (10 µm thickness). **b, d** Complex quartz response function $h_Q$ for 50 µm is experimentally extracted using known ZnTe response (blue) and modeled (red) in amplitude and phase. **c, e** Calculated $\chi^{(2)}$, transmitted field coefficient $t_F$, and phase matching factor $G$ in amplitude and phase, showing how these factors contribute to the quartz response function.

$$h(\Omega) = \chi_{\text{eff}}^{(2)}(\Omega) t_F(\Omega) \int_{\omega > \Omega} d\omega \, \frac{\omega^2}{c^2 k(\omega)} T_s(\omega, \Omega) E_s^*(\omega) E_s(\omega - \Omega) \, G(\omega, \Omega), \qquad (1)$$

where $h_{\text{calc}}(\Omega) = [h(\Omega) + h^*(-\Omega)]/(I_1 + I_2)$ and $I_1 + I_2 = \int d\omega \, T_s(\omega, \Omega = 0) E_s^*(\omega) E_s(\omega)$. Here, $E_s$ is the incident sampling pulse with optical frequency $\omega$ and wavenumber $k(\omega) = n(\omega)\omega/c$, where $n(\omega)$ is the corresponding refractive index. $T_s(\omega, \Omega)$ accounts for the sampling pulse transmission $T_s(\omega, \Omega) = t_{12}^*(\omega) t_{21}^*(\omega) t_{12}(\omega - \Omega) t_{21}(\omega - \Omega)$, where $t_{12}(\omega)$ and $t_{21}(\omega)$ are the Fresnel transmission coefficients for propagating from air into quartz and quartz into air, respectively. $\chi_{\text{eff}}^{(2)}(\Omega) = \chi_{\text{eff}}^{(2)}(\omega_c, \Omega)$ is the effective nonlinear susceptibility of the detection crystal under the assumption that $\chi_{\text{eff}}^{(2)}(\omega - \Omega, \Omega) \approx \chi_{\text{eff}}^{(2)}(\omega_c, \Omega)$, where $\omega_c$ is the sampling-pulse center frequency. The field transmission coefficient $t_F(\Omega)$ accounts for the transmitted THz field including its multiple reflections (see **Methods**). The phase-matching factor, $G(\omega, \Omega) = [\exp(i\Delta k(\omega, \Omega)d) - 1]/i\Delta k(\omega, \Omega)$ between THz and sampling pulse includes $\Delta k(\omega, \Omega) = k(\omega - \Omega) + k(\Omega) - k(\omega)$ and the sample thickness $d$.

To calculate $h_Q$, we use the known quartz refractive indices in the THz[14] and optical region[45]. However, the nonlinear susceptibility $\chi^{(2)}(\Omega)$ is not known and we therefore model it by:

$$\chi_{\text{eff}}^{(2)}(\Omega) = \chi_e^{(2)} \left[ 1 + B(1 - i\Omega\tau_D)^{-1} + C \left( 1 - \frac{\Omega^2}{\Omega_{\text{TO}}^2} - \frac{i\Omega\Gamma}{\Omega_{\text{TO}}^2} \right)^{-1} \right], \qquad (2)$$

where $\chi_e^{(2)}$ is the pure electronic susceptibility. The last term corresponds to the ionic contribution with $\omega_{TO}$ being the frequency, and $\Gamma$ being the damping of the respective transverse-optical (TO) phonon, while the Faust-Henry coefficient $C$ defines the ratio between the lattice-induced and electronic contributions[29,46]. We take the phonon parameters $\Omega_{TO}/(2\pi) = 3.9$ THz, and $\Gamma/(2\pi) = 0.09$ THz from Davies *et al.*[14] and find $C = 0.15$ to provide good agreement with our experimental values (see red curves in **Figs. 2b,d**). We assume that the striking low-frequency enhancement of $h_Q(\Omega)$ (see **Fig. 2b**) arises from $\chi^{(2)}$ and model it by a phenomenological Debye-type relaxation contribution $B$ with characteristic time scale $\tau_D$ (second term in **Eq. (2)**). Choosing $B = 0.7$ and $\tau_D = 0.5$ ps provides nearly perfect agreement with the 0.1-0.9 THz range in $h_{Q,exp}$. We will discuss possible physical origins of such a contribution below. Thus, by analytic modeling, we find dominating contributions by the phase matching factor $G$, the field transmission coeffient $t_F$, and the nonlinear susceptibility $\chi^{(2)}$, disentangled in **Fig. 2c,e**.

We apply the response function to calculate the exact THz electric field (red) from the quantitative EOS signal in 50 µm quartz (blue) in **Figs. 3a** and **3b** in the time- and frequency-domain, respectively. To determine the absolute field strength, we use the measured THz pulse energy and focal size (see **Supplementary Note 1**). We obtain a peak field strength of 1.04 MV/cm. We can therefore estimate the effective electrooptic coefficient $r_{eff}$, which equals the $r_{11}$ tensor component, of z-cut quartz to be 0.1 pm/V (see **Supplementary Note 2**). This value agrees well with previous reports of $r_{11}$ at optical frequencies ranging between 0.1 and 0.3 pm/V in z-cut quartz[43,47].

*Thickness Dependence and Nonlinear Origin*

The response function also depends on the crystal thickness, which typically presents a trade-off between sensitivity and bandwidth. **Fig. 3c** shows the measured dependence of the maximum EOS signal on the quartz crystal thickness between 35 and 150 µm (blue dots), which clearly deviates from an ideal phase-matched behavior, i.e., a linear scaling with the crystal thickness. We also observe a noticeable thickness dependence of the time-domain EOS shapes in **Fig. 3d**, even clearer in the spectral bandwidth in **Fig. 3e**. **Fig. 3f** displays the calculated response function for each thickness in amplitude (red) and phase (grey), which explains the measured features. For instance, the effective bandwidth is significantly lower for 150 µm quartz due to the zero in the phase-matching factor $G(\Omega, \omega)$, while the thickness-dependent frequency spacing of the modulations generally arise from Fabry-Perot fringes in the field transmission coefficient $t_F(\Omega)$. The calculated response function, thus, also explains the experimentally observed EOS thickness dependence in **Fig. 3c** (red line), mainly by the phase mismatch $G(\Omega, \omega)$ of THz and sampling pulse.

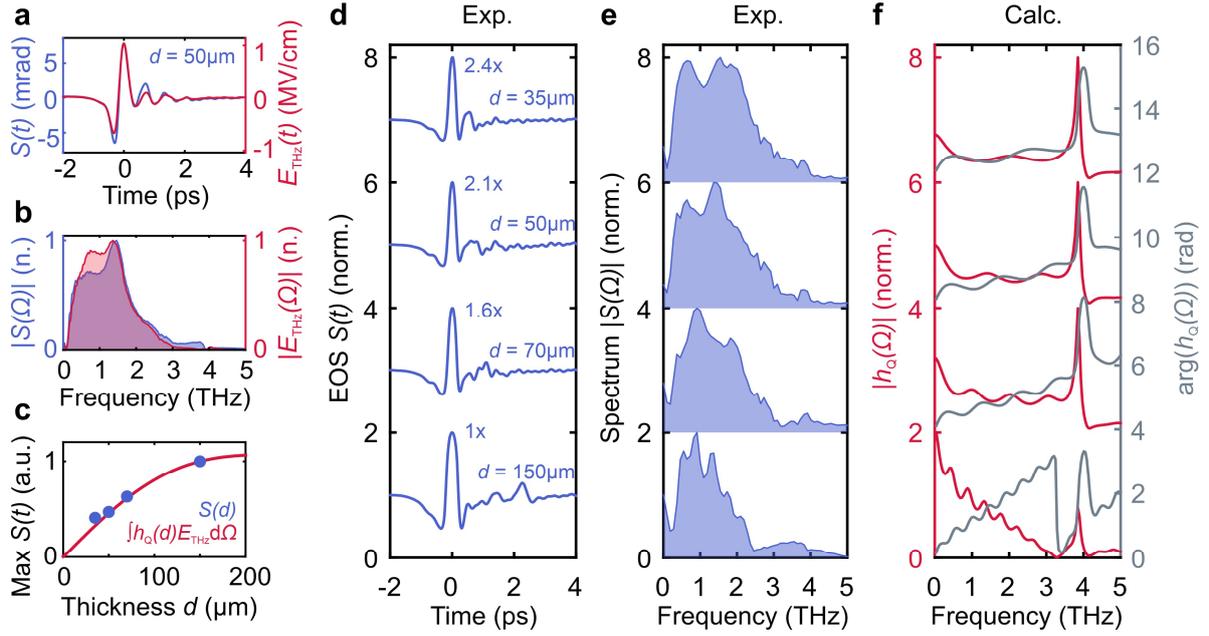

**Fig. 3 | Thickness dependence and extracted THz electric fields. a** Absolute THz electric field $E_{THz}$ extracted by applying the response function $h_Q$ to the measured EOS signal $S$ with 50 μm quartz and **b** corresponding Fourier amplitude spectrum. **c** Maximum EOS signal $S(t)$ as a function of quartz thickness (blue markers) and calculated quartz response (red curve). **d** EOS signal for four different quartz thicknesses below 150 μm with **e** respective Fourier amplitude spectra. **f** Modulus and phase of calculated detector response $h_Q$ of quartz for the respective thicknesses. The small oscillatory variations below 4 THz are Fabry-Perot resonances. The zero in (e) for 150 μm is dictated by the phase matching factor $G$. The peak at 3.9 THz stems from the phonon contribution to $\chi^{(2)}$.

The first report of EOS in quartz suggested a strong surface $\chi^{(2)}$ contribution[42]. Indeed, the surface and bulk $\chi^{(2)}$ have a similar order of magnitude[48]. As the surface contribution originates from a depth of ~1 nm (Ref. 48), its contribution will be small in comparison to the bulk contribution for a quartz crystal with a thickness >10 μm. The response functions presented here (**Figs. 2b,d** and **3f**) strongly indicate a pure bulk $\chi^{(2)}$ effect and provide a reasonable estimate of $r_{11}$, both sufficient to explain the experimental observations.

We suggest the low-frequency (0.1-0.9 THz) enhancement in $\chi^{(2)}$ to be caused by disorder. In fact, the frequency region 0.1-1.2 THz of fused silica and other glasses is often associated with the so-called Boson-peak behavior corresponding to low frequency vibrational modes[49,50]. Its nature and origin remain debated, but it is known to affect the Raman, neutron, and linear dielectric responses of quartz and related glasses[49-51]. Our finding, thus, motivates further research into the nonlinear susceptibility in the sub-0.9 THz region. In addition, there is considerable variability of the reported values for the 3.9 THz phonon damping parameter $\Gamma/2\pi$ between 0.09 THz (Ref. 14) and 0.39 THz (Ref. 51). This variation indicates that the $\chi^{(2)}$ model

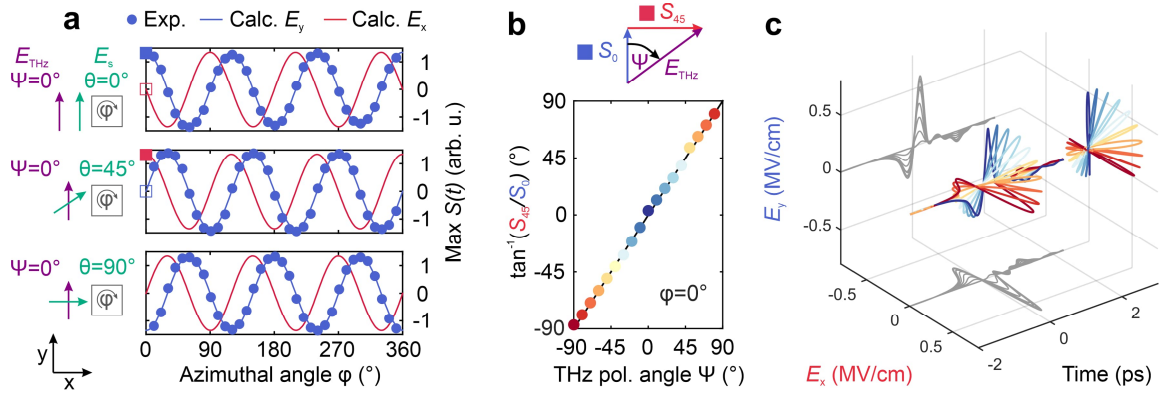

**Fig. 4 | Polarization and azimuthal angle dependence for 2D-EOS. a** Measured azimuthal angle $\phi$ dependence of maximum quartz $S(t)$ for different sampling pulse polarizations ($\theta$) with THz pulse polarized along y (blue dots). Blue and red lines are the calculated azimuthal angle dependence for the respective sampling pulse polarizations and THz polarized along y (blue line) and x (red line). **b** The arctan of the peak EOS signals measured at $\theta = 45°$ ($S_{45}$) and $\theta = 0°$ ($S_0$) perfectly matches the THz polarizer angle $\psi$, demonstrating that the full THz polarization state can be extracted by measuring $S_0(t)$ and $S_{45}(t)$. **c** 2D-EOS: $E_x^{\mathrm{THz}}(t)$ and $E_y^{\mathrm{THz}}(t)$ for selected $\psi$ between 0° and 90°, which were extracted from $S_{45}(t)$ and $S_0(t)$ by applying the quartz response function $h_{\mathrm{Q}}$.

parameters are highly sensitive to the sample quality and may be fine-tuned for better agreement.

### *Polarization-Resolved EOS*

So far, we have treated both $h_{\mathrm{Q}}$ and $E_{\mathrm{THz}}$ as scalars and only considered the specific case in which the THz-pulse and sampling-pulse polarizations are parallel and the quartz azimuthal angle is optimized for maximum $S(t)$, i.e., oriented parallel to one of the in-plane crystalline axes. However, the THz electric field is a vectorial observable and can have an arbitrary (and thus even helical) polarization state and $h_{\mathrm{Q}}$ is generally dependent on the azimuthal angle $\phi$ and sampling pulse polarization $\theta$. Nonetheless, we can assume the same frequency-dependence of the allowed $\chi^{(2)}$ tensor elements and any corresponding linear combination of them, because of the in-plane symmetry of the 3.9 THz phonon. Since the other quantities in **Eq. (1)**, such as $t_{\mathrm{F}}$ or $G$, refer to linear optical properties, they are also in-plane isotropic in z-cut quartz. We can, therefore, assume the same frequency evolution of the response function for all $\phi$ and $\theta$, but the absolute sensitivity will be rescaled by the global symmetry of $\chi_e^{(2)}(\phi, \theta)$, which ultimately allows for polarization-sensitive THz EOS.

To explore the sensitivity of 50 μm quartz to different THz field polarization components, **Fig. 4a** shows the measured peak EOS signal $S$ (blue dots) as a function of quartz azimuthal angle $\phi$ for three different probe polarizations $\theta = 0°, 45°, 90°$ with respect to the THz field ($\psi = 0°$, linearly polarized along the y-axis). Each azimuthal dependence $S(\phi)$ exhibits a perfect 3-fold symmetry in agreement with the first reported quartz EOS[42]. We therefore calculate the expected dependence of $S(\psi, \theta)$ for a THz field $\mathbf{E}_{\text{THz}}$ linearly polarized at an arbitrary angle $\psi$, and sampling field $\mathbf{E}_s$ linearly polarized at angle $\theta$ in the x-y plane (see **Fig. 1a**). We use the 2nd order nonlinear polarization $P_i^{(2)} = \epsilon_0 \chi_{ijk}^{(2)} E_j^{\text{THz}} E_k^s$, which we can rewrite using the nonlinear susceptibility tensor in contracted notation $d_{il}$ with only non-zero $d_{11}$ and $d_{14}$ terms due to quartz's $D_3$ point group, evaluated for the z-cut plane[52] (see **Methods**). The blue line in **Fig. 4a** shows the expected sensitivity for a vertically polarized THz field $E_y^{\text{THz}}$ (i.e. $\psi = 0°$), in perfect agreement with the measured azimuthal dependence. The expected peak signal for a horizontally polarized THz field $E_x^{\text{THz}}$ (i.e. $\psi = 90°$), shown as a red line, features the same 3-fold symmetry but shifted by 30°. These opposite EOS sensitivities for the x- and y-projections of the THz field allow for a full THz polarization determination by simply measuring EOS for two different sampling pulse polarizations, e.g. $\theta = 0°$ for obtaining $E_y^{\text{THz}}$ and $\theta = 45°$ for obtaining $E_x^{\text{THz}}$ at azimuth $\phi = 0$ (see square markers in **Fig. 4a**).

To prove this concept, we rotate the linear polarization of the THz pulse by setting polarizer P1 to 45° and scanning P2 by angle $\psi$. Next, we measure $S(t)$ for sampling pulse polarization $\theta = 0°$ ($S_0$) and 45° ($S_{45}$) for a set of THz polarizer angles $\psi$. **Fig. 4b** shows that $\arctan(S_{45}/S_0)$ is identical to the THz polarizer angle $\psi$ and, thus, precisely measures the THz polarization by only two EOS measurements at different sampling pulse polarizations. After applying the calculated response function $h_Q$ to $S_0$ and $S_{45}$, the full vectorial THz field $\mathbf{E}_{\text{THz}}(t)$ can be extracted as shown in the 2D EOS traces for selected $\psi$ between 0° and 90° in **Fig. 4c**. We note that the perfect 3-fold symmetry is not found in the common ZnTe (110) or GaP (110) EOS crystals, where this convenient procedure cannot be used[36].

*Broadband THz Helicity Measurement*

For driving chiral or, generally, helicity-dependent excitations, e.g., for ultrafast control of phonon angular momentum[20,21,53] or topology modulation[18], CEP-stable table-top THz sources are beneficial due to their inherent synchronization with sub-cycle probing pulses. Nevertheless, to reach the required peak fields, the energy has to be squeezed into few- or single cycle pulses at low repetition rates. Therefore, the lack of broadband THz waveplates leads to complicated polarization states when aiming for THz pulses with specific helicities. In contrast to conventional multi-cycle optical light, helical few- or single-cycle THz pulses are highly polychromatic and, generally, cannot be described by a single polarization state, i.e., neither by a pair of ellipticity angles $(\vartheta, \eta)$ nor by one fixed Jones or Stokes vector[44]. Instead, the polarization state must be generally described as an evolution in frequency space or, equivalently, by the full temporal trajectory of the light's electric field vector $\mathbf{E}_{\mathrm{THz}}(t)$.

To demonstrate the complete detection of arbitrary polarization states in quartz, in particular for complicated helical fields, we characterize the polarization state of single-cycle THz pulses following collimated traversal of the textbook birefringent y-cut quartz (see **Supplementary Fig. S1**), which is nearly identical to commercially available THz waveplates. **Fig. 5a** shows the transmitted electric field of a collimated THz beam ($\psi = 45°$) through 0.7 mm crystalline y-cut quartz for three different crystal orientations, which is detected in 50 μm z-cut quartz. The transmitted THz polarizations for 0° and 90° orientations appear highly elliptical, which is when the incident THz pulse polarization is at 45° to the in-plane crystal axes and therefore experiences maximum birefringence. This form of time-domain ellipsometry permits the direct measurement of the birefringence $\Delta n(\Omega)$ using $\arg(E_x^{\mathrm{THz}}) - \arg(E_y^{\mathrm{THz}}) = \Delta n \Omega d/c$ as shown in **Fig. 5b**. We find an approximately constant $\Delta n(\Omega)$ of about 0.05 at 0.4-3.5 THz, in good agreement with literature values[51,54].

As seen from **Fig. 5a**, the transmitted THz polarization is neither a simple polarization ellipse nor purely left- or right-handed circularly polarized. Its sophisticated electric-field trajectory can be described by a frequency-dependent rotation $\vartheta(\Omega)$ and ellipticity $\eta(\Omega)$ (see inset **Fig. 5a**), or any other ellipsometric set of parameters as a function of frequency. **Figs. 5c-e** show $\vartheta(\Omega)$ and $\eta(\Omega)$ for 0°, 45°, and 90° orientation of the y-cut quartz plate, respectively. For 0° and 90° orientation (see **Figs. 5c,e**), the THz pulse acquires a maximum of frequency-dependent ellipticity $\Delta n \Omega d/c$. Since $\Delta n(\Omega)$ is roughly constant, the transmitted THz pulse for 0° and 90° orientation is, respectively, perfectly right- and left-handed circularly polarized only at frequency $c/(4\Delta n d) \approx 2.37$ THz and 1.96 THz, where $\eta$ reaches -45° and 45°. In other words, the y-cut quartz plate acts as a THz quarter waveplate (QWP) for only a very narrow frequency range and leads to drastically different polarization states for all other frequency components within the THz pulse (see top row of **Figs. 5c,e**). In contrast, the incident THz pulse acquires a small ellipticity for the 45° orientation (see **Fig. 5d**) only at higher frequencies, which are more sensitive to a small $\Delta n$.

Usually, broadband QWPs create opposite helicities for ±45° rotation. This behavior is evidently not the case here, as the two $\mathbf{E}(t)$ trajectories in **Fig. 5f** are not perfectly opposite. We project the polarization state from a linear into a circular basis to resolve the frequency-dependent helicity (see **Methods**). **Figs. 5g,h** depict the full frequency-dependent right-handed ($E_{\mathrm{RCP}}$) and left-handed ($E_{\mathrm{LCP}}$) circularly polarized intensity components for the 0° (red) and 90° (blue) orientations, normalized for every frequency component (see **Fig. 5g**) and as absolute intensity spectra (see **Fig. 5h**). **Fig. 5g** highlights that the helicity changes quite drastically across the single THz pulse spectrum and that a circular polarization is achieved at slightly different frequencies for opposite QWP angles (0° vs. 90°), in agreement with the ellipticity parameters $\eta(\Omega)$ in **Figs. 5c,e**. The latter can be related to a slightly tilted axis of rotation with respect to the quartz plate's y-axis, which highlights the challenges of helicity-dependent measurements in the THz spectral range.

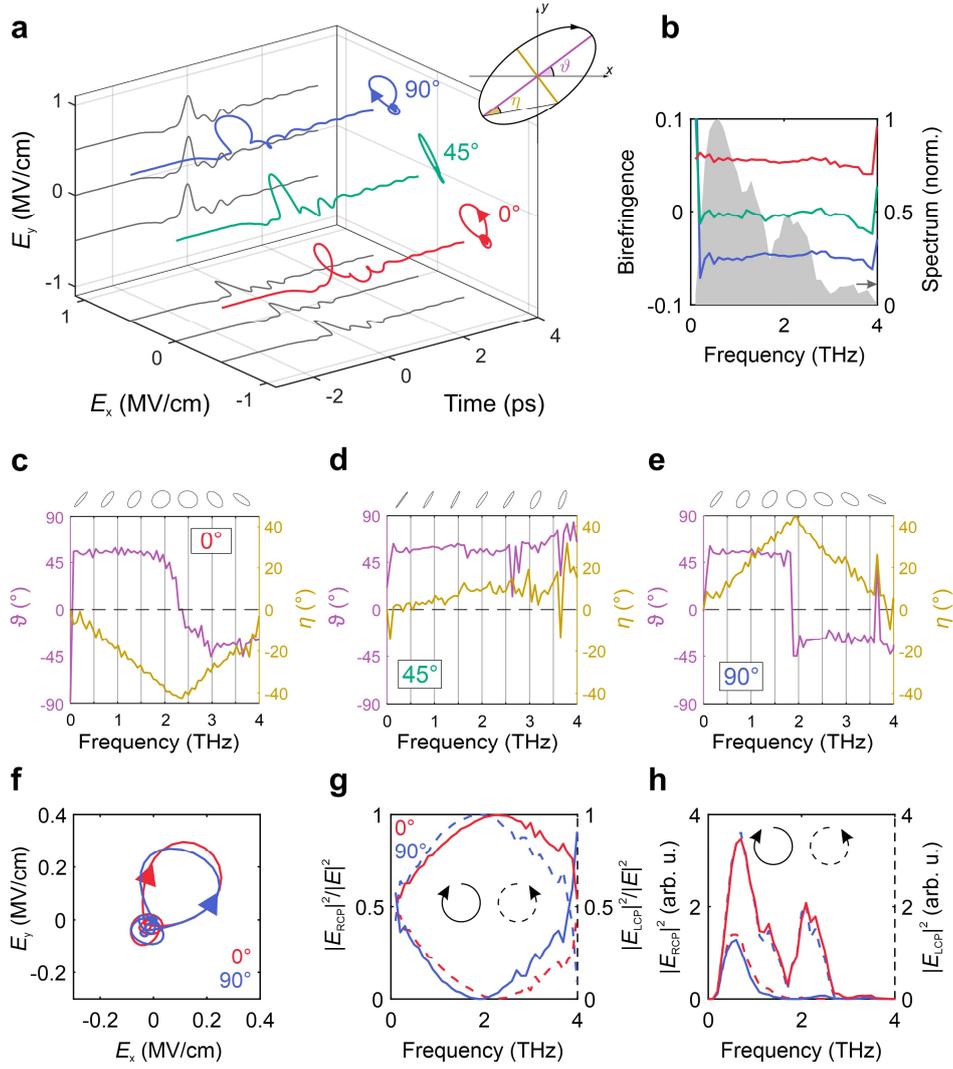

**Fig. 5 | Detection of arbitrary THz polarization states and their helicity. a** 2D-EOS of the THz electric field transmitted through a 0.7 mm y-cut quartz plate for three different y-cut quartz orientations, detected in 50 µm z-cut quartz . The y-cut quartz plate was aligned with one of its facets parallel to the y-axis (corresponds to 0°). The incident THz field was linearly polarized at 45°. **b** Extracted birefringence for the three different y-cut quartz azimuthal angles, demonstrating that the THz field experiences the largest birefringence for 0° and 90° quartz-plate orientations. **c, d, e** Corresponding frequency-resolved THz polarization states expressed in polarization ellipse rotation $\vartheta(\Omega)$ and ellipticity $\eta(\Omega)$ for 0°, 45° and 90° y-cut plate azimuthal angle, respectively. **f** Projection of $E_x^{\mathrm{THz}}(t)$ and $E_y^{\mathrm{THz}}(t)$ into the $(E_x^{\mathrm{THz}}, E_y^{\mathrm{THz}})$ plane for 0° and 90° y-cut plate azimuthal angles, unveiling the different, but not exactly opposite helicity states. **g** Corresponding LCP and RCP intensity spectra normalized for every frequency $\Omega$ to $|E^{\mathrm{THz}}(\Omega)|^2$ and **h** corresponding absolute intensities.

**Discussion**

We now discuss the detector performance of α-quartz in more detail. As we find a pure bulk $\chi^{(2)}$ effect, the phase-matching term $G$ governs the trade-off between detection sensitivity and bandwidth. The effective detector bandwidth is, thus, limited by the first zero in $G$ (**Fig. 3f**), giving a cut-off frequency $\nu_{\mathrm{cuttoff}} = c / \left[ \left( n_{\mathrm{THz}} - n_{\mathrm{s}}^{(g)} \right) d \right] = 1 / (\mathrm{GVM} \cdot d)$. The group-velocity mismatch (GVM) in quartz is about 1.8 ps/mm (assuming $n_{\mathrm{THz}}$(1 THz) = 2.09 and group index $n_{\mathrm{s}}^{(g)}$(800 nm) = 1.55), which is only slightly inferior to ZnTe (GVM = 1.1 ps/mm)[30]. A full comparison of $r_{\mathrm{eff}}$ and GVM between quartz and the widely used EOS crystals ZnTe and GaP is shown in **Supplementary Table S1**. Therefore, to sample the whole THz spectrum of typical high-field THz sources based on LiNbO$_3$ (~0.1-4 THz), the quartz thickness should not exceed 130 µm. Sampling of higher THz frequencies poses limitations due to substantial dispersion of the linear THz refractive index and nonlinear susceptibility $\chi^{(2)}$ due to the 3.9, 8, 12, and 13.5 THz TO phonons of quartz[55]. This fact is especially relevant for more broadband high-field sources such as large-area spintronic emitters[3,42].

The polarization sensitivity of EOS in quartz generally permits time-domain ellipsometry, allows for the direct measurement of complex and even non-equilibrium[24] tensorial material properties in anisotropic media[22,23] and optical activity of chiral phonons[25,26,56], as well as THz circular-dichroism spectroscopy[26,27], or decoding high-harmonic THz emission of complex quantum materials[15-17]. The ability to detect intense THz fields in amplitude and phase without distortions is well suited for any ultrafast spectroscopy based on strong THz-field excitation[13], e.g., for understanding nonlinear THz polarization responses[38] or driving phase transitions[8-10,18], where an accurate characterization of the driving field is crucial. Moreover, the demonstrated precise helicity characterization of intense THz driving fields is urgently needed for the emerging field of chiral (or circular) phononics. In this field, lattice modes are driven on chiral or circular trajectories with phonon angular momentum[53] leading to magnetization switching[57], transient multiferroicity[21], large magnetic fields[20] or other yet unexplored spin-lattice-coupled phenomena. These first explorations in the uncharted territory of phonon-angular-momentum control highlight the challenges for THz helicity differential detection, i.e., extracting signals proportional to $S(E_{\mathrm{RCP}}) - S(E_{\mathrm{LCP}})$, which must be employed to isolate helicity-dependent effects. Using quartz as a reliable high-field THz helicity detector will help to clarify and support these novel types of measurements and will foster further studies of chiral or helicity-selective phenomena in the THz spectral region.

As the demonstrated 2D-EOS protocol only relies on a single HWP rotation, it enables a rapid measurements and, therefore, keeps the phase error due to temporal drifts between adjacent EOS scans minimal. Accordingly, the scheme is also easy to implement in commercial time-domain spectrometer systems as it only relies on the addition of low-cost and widely available thin quartz wafers and standard HWPs in the VIS or NIR spectral range. As another benefit, quartz is well suited for measuring THz fields and their polarization states in systems, where space constraints often prohibit the use of motorized rotation mounts for the detection crystal, in particular in cryostats at cryogenic temperatures. **Supplementary Fig. S2** shows quartz EOS at 80 K, demonstrating that the THz field can still be reliably sampled at low temperatures, although the response function is modified due to the enhanced phonon contribution to $\chi^{(2)}$ (Ref. 14). Conveniently, our work may also allow for all-optical synchronization of THz pump and optical probe pulses via THz slicing[58] or in-situ field and polarization characterization in already installed z-cut quartz windows at free-electron-laser facilities, where even a non-collinear THz- and sampling-beam geometry is feasible (see **Supplementary Fig. S4**).

In conclusion, z-cut α-quartz can reliably sample intense THz fields of the order of 1 MV/cm without over-rotation and with negligible higher-order nonlinearities. We measured and modeled the frequency-dependent electrooptic response function, consistent with a pure bulk $\chi^{(2)}$ effect dominated by Fabry-Perot resonances, phonon modulations in the Faust-Henry formalism, phase matching effects, and a low frequency Debye-like contribution. We determined the electrooptic coefficient to the order of 0.1 pm/V and proved a perfect 3-fold symmetry of the electrooptic response. Based on this knowledge, we developed an easily implementable protocol to measure the full vectorial THz polarization state by simply toggling between 0° and 45° sampling pulse polarizations. With this approach, we establish quartz as a powerful detector for full amplitude, phase and polarization state of highly intense THz radiation at a fraction of the cost of conventional detection crystals. This work will accordingly foster rapid and cost-efficient high-field THz spectroscopy[5,6,13,15], THz time-domain ellipsometry[23], THz circular-dichroism spectroscopy[26,27] and will enable broadband THz helicity characterization of polarization-tailored pulses for driving angular-momentum phonons[25,26,56] or other helicity-dependent excitations[18-21] in the future.

**Methods**

*Generation and electrooptic sampling of intense THz pulses*

Intense THz pulses (1.3 THz center frequency, 1.5 THz FWHM) with peak fields exceeding 1 MV/cm are generated using optical rectification in LiNbO$_3$ using the titled pulse front technique[1]. For this, the LiNbO$_3$ crystal is pumped with laser pulses from an amplified Ti:sapphire laser system (central wavelength 800 nm, pulse duration 35 fs FWHM, pulse energy 5 mJ, repetition rate 1 kHz). The THz field strengths are altered by rotating THz polarizer P1 (Fig. 1a), while keeping P2 fixed at $\psi = 0°$. In this way, the peak fields of the transmitted THz pulses are proportional to the cosine squared of the polarizer P1 angle. Similarly, the THz polarization $\psi$ can be set to an arbitrary angle by keeping P1 fixed at 45° and P2 at $\psi$. The sampling pulses are provided by a synchronized Ti:sapphire oscillator (central wavelength 800 nm, repetition rate 80 MHz) and are collinearly aligned and temporarily delayed with respect to the THz pulse. The sampling pulse polarization is set to specific angles by using a half-waveplate (HWP) before the EOS crystal.

The THz pulse induces a change in birefringence (electrooptic effect or Pockels effect) in the EOS crystal. This birefringence causes the sampling pulse to acquire a phase difference between its polarization components parallel and perpendicular to the THz pulse polarization. This phase difference is detected in a balanced detection scheme consisting of a quarter- and half-waveplate ($\lambda/4$, $\lambda/2$) followed by a Wollaston prism (WP) to spatially separate the perpendicular polarization components. The intensity of the two resulting beams is detected by two photodiodes ($I_1$ and $I_2$), which leads to the EOS signal $S = (I_1 - I_2)/(I_1 + I_2)$, that is equal to twice the THz-induced phase difference (see **Supplementary Note 2**).

*Further details of the electrooptic response function model*

The electrooptic response is modeled using **Eq. (1)** and **(2)** in the main text. In this equation, the field transmission coefficient $t_F(\Omega)$ accounts for the transmitted THz field, includes multiple reflections of the field inside the crystal, and can be expressed using:

$$t_F(\Omega) = \frac{t_{12}(\Omega)}{1 - R(\Omega)e^{i\theta(\Omega)}},\qquad(3)$$

where $R(\Omega) = r_{21}(\Omega)r_{23}(\Omega)$, and $\theta = 2\Omega n(\Omega)d/c$. Here, $t_{12}(\Omega)$ and $r_{12}(\Omega)$ are the Fresnel transmission and reflection coefficients at THz frequencies at the respective interfaces (1,3 - air; 2 - quartz), respectively.

*EOS response for arbitrary quartz azimuthal angles and sampling pulse polarizations*

To compute the full quartz EOS dependence for the crystalline azimuthal angle $\phi$, sampling pulse polarization $\theta$, and THz polarization angle $\psi$, we consider the second-order nonlinear polarization $\mathbf{P}^{(2)}$, which for z-cut α-quartz ($D_3$ point group[52]) can be written using contracted notation in the matrix form:

$$
\begin{bmatrix} P_x^{(2)} \\ P_y^{(2)} \\ P_z^{(2)} \end{bmatrix} = 4\epsilon_0 \begin{bmatrix} d_{11} & -d_{11} & 0 & d_{14} & 0 & 0 \\ 0 & 0 & 0 & 0 & -d_{14} & -d_{11} \\ 0 & 0 & 0 & 0 & 0 & 0 \end{bmatrix} \begin{bmatrix} E_x^s E_x^{\mathrm{THz}} \\ E_y^s E_y^{\mathrm{THz}} \\ E_z^s E_z^{\mathrm{THz}} \\ E_y^s E_z^{\mathrm{THz}} + E_z^s E_y^{\mathrm{THz}} \\ E_x^s E_z^{\mathrm{THz}} + E_z^s E_x^{\mathrm{THz}} \\ E_x^s E_y^{\mathrm{THz}} + E_y^s E_x^{\mathrm{THz}} \end{bmatrix}, \tag{4}
$$

where $d_{11} = 0.3$ pm/V and $d_{14} = 0.008$ pm/V (Ref. 52). For our experimental configuration, the probe and THz polarizations are in the x-y plane, and the z components of these fields are zero. **Equation (4)** thus implies that only the $d_{11}$ component affects quartz EOS in our geometry. The balanced-detection signal is proportional to the difference of the intensities of the orthogonally polarized x- and y-components of the total electric field at the detector, separated by the Wollaston prism and projected on the two photodiodes. Thus, the signal can be calculated as:

$$
S \propto I_1 - I_2 \approx (E_x^s + E_x^{(2)})^2 - (E_y^s + E_y^{(2)})^2, \tag{5}
$$

where $\mathbf{E}^{(2)}$ is the electric field emitted by the nonlinear polarization $\mathbf{P}^{(2)}$ as described by the inhomogeneous wave equation. The sampling-pulse polarization angle is defined by $\theta = \mathrm{atan2}(E_x^s, E_y^s)$ and the THz polarization angle by $\psi = \mathrm{atan2}(E_x^s, E_y^s)$, where $\mathrm{atan2}$ corresponds to the four-quadrant arctan function.

A convenient way to numerically simulate the nonlinear polarization $\mathbf{P}^{(2)}$ obtained with an azimuthal rotation of the sample in the x-y plane by an angle $\phi$ is to apply two-dimensional rotation matrices $R(\phi)$ to the $\mathbf{E}_s$ and $\mathbf{E}_{\mathrm{THz}}$ fields while using an unchanged form of $d_{ij}$ in **Eq. (4)** and then rotate the calculated nonlinear polarization $\mathbf{P}^{(2)\prime}$ by $-\phi$ back into the original lab frame. After the rotation by $R(\phi)$, the sampling and THz field components take the form $\mathbf{E}_s'(\phi) = R(\phi)\mathbf{E}_s$ and $\mathbf{E}_{\mathrm{THz}}'(\phi) = R(\phi)\mathbf{E}_{\mathrm{THz}}$, and allow $\mathbf{P}^{(2)\prime}(\phi)$ to be computed. Rotating back to the lab frame, then yields $\mathbf{P}^{(2)}(\phi) = R(-\phi)\mathbf{P}^{(2)\prime}(\phi)$. The signal azimuthal angle dependence $S(\phi)$ can then be calculated using **Eq. (5)** as before. Since $\theta$ and $\psi$ can be set arbitrarily, the full $S(\phi, \theta, \psi)$ dependence of quartz can be constructed (see **Supplementary Fig. S3**).

*Polarization state representations of polychromatic THz fields*

The polarization state of a THz field $\mathbf{E}(\Omega)$ can be described using a polarization ellipse representation (see inset **Fig. 5a**), where the orientation $\vartheta(\Omega)$ is given by:

$$
\vartheta(\Omega) = \frac{1}{2}\arctan\left(\frac{2|E_x(\Omega)||E_y(\Omega)|}{|E_x(\Omega)|^2 - |E_y(\Omega)|^2}\cos(\Delta(\Omega))\right), \tag{6}
$$

with $\Delta(\Omega) = \arg(E_y(\Omega)) - \arg(E_x(\Omega))$. The ellipticity $\eta(\Omega)$ is given by

$$\eta(\Omega) \frac{1}{2} \arcsin \left( \frac{2|E_x(\Omega)||E_y(\Omega)|}{|E_x(\Omega)|^2 + |E_y(\Omega)|^2} \sin(\Delta(\Omega)) \right). \qquad (7)$$

Another useful way to describe $\mathbf{E}(\Omega)$, which is typically measured in a linear basis $(\hat{\mathbf{x}}, \hat{\mathbf{y}})$, is the circular basis $(\hat{\mathbf{R}}, \hat{\mathbf{L}})$. In this representation, the right- and left-hand circular polarized field components, $E_{\mathrm{RCP}}(\Omega)$ and $E_{\mathrm{LCP}}(\Omega)$ respectively, are given by the projection:

$$E_{\mathrm{RCP}}(\Omega) = \frac{1}{\sqrt{2}}[E_x(\Omega) + iE_y(\Omega)], \qquad (8)$$

$$E_{\mathrm{LCP}}(\Omega) = \frac{1}{\sqrt{2}}[E_x(\Omega) - iE_y(\Omega)]. \qquad (9)$$

**Acknowledgements:**

We thank A. Paarmann, Y. Behovits, A. Chekhov, and M. Wolf for fruitful discussions. **Funding:** This project was mainly funded through S.F.M.'s Emmy Noether Independent Junior Research Group from the Deutsche Forschungsgemeinschaft (DFG, German Research Foundation, no. 469405347).

**Supplementary information:**

Supplementary information accompanies this manuscript.

**References:**

1. Hirori H, Doi A, Blanchard F, Tanaka K. Single-cycle terahertz pulses with amplitudes exceeding 1 MV/cm generated by optical rectification in LiNbO3. *Appl Phys Lett* **98**, (2011).
2. Sell A, Leitenstorfer A, Huber R. Phase-locked generation and field-resolved detection of widely tunable terahertz pulses with amplitudes exceeding 100 MV/cm. *Optics Letters* **33**, 2767-2769 (2008).
3. Rouzegar R, *et al.* Broadband Spintronic Terahertz Source with Peak Electric Fields Exceeding 1.5 MV/cm. *Physical Review Applied* **19**, (2023).
4. Green B, *et al.* High-Field High-Repetition-Rate Sources for the Coherent THz Control of Matter. *Sci Rep* **6**, 22256 (2016).
5. Maehrlein S, Paarmann A, Wolf M, Kampfrath T. Terahertz Sum-Frequency Excitation of a Raman-Active Phonon. *Phys Rev Lett* **119**, 127402 (2017).
6. Johnson CL, Knighton BE, Johnson JA. Distinguishing Nonlinear Terahertz Excitation Pathways with Two-Dimensional Spectroscopy. *Phys Rev Lett* **122**, 073901 (2019).
7. Kampfrath T, *et al.* Coherent terahertz control of antiferromagnetic spin waves. *Nature Photonics* **5**, 31-34 (2010).
8. Fausti D, *et al.* Light-Induced Superconductivity in a Stripe-Ordered Cuprate. *Science* **331**, 189-191 (2011).
9. Li X, *et al.* Terahertz field-induced ferroelectricity in quantum paraelectric SrTiO3. *Science* **364**, 1079-1082 (2019).
10. Disa AS, *et al.* Photo-induced high-temperature ferromagnetism in YTiO(3). *Nature* **617**, 73-78 (2023).
11. Maehrlein SF, *et al.* Dissecting spin-phonon equilibration in ferrimagnetic insulators by ultrafast lattice excitation. *Sci Adv* **4**, (2018).

12. Bell G, Hilke M. Polarization Effects of Electro-optic Sampling and Over-rotation for High Field THz Detection. *Journal of Infrared, Millimeter, and Terahertz Waves* **41**, 880-893 (2020).

13. Leitenstorfer A*, et al.* The 2023 terahertz science and technology roadmap. *Journal of Physics D: Applied Physics* **56**, (2023).

14. Davies CL, Patel JB, Xia CQ, Herz LM, Johnston MB. Temperature-Dependent Refractive Index of Quartz at Terahertz Frequencies. *Journal of Infrared, Millimeter, and Terahertz Waves* **39**, 1236-1248 (2018).

15. Tielrooij KJ*, et al.* Milliwatt terahertz harmonic generation from topological insulator metamaterials. *Light Sci Appl* **11**, 315 (2022).

16. Hafez HA*, et al.* Extremely efficient terahertz high-harmonic generation in graphene by hot Dirac fermions. *Nature* **561**, 507-511 (2018).

17. Chu H*, et al.* Phase-resolved Higgs response in superconducting cuprates. *Nat Commun* **11**, 1793 (2020).

18. Sie EJ*, et al.* An ultrafast symmetry switch in a Weyl semimetal. *Nature* **565**, 61-66 (2019).

19. Ueda H*, et al.* Chiral phonons in quartz probed by X-rays. *Nature* **618**, 946-950 (2023).

20. Luo J*, et al.* Large effective magnetic fields from chiral phonons in rare-earth halides. *arXiv:230603852*, (2023).

21. Basini M*, et al.* Terahertz electric-field driven dynamical multiferroicity in SrTiO3. *arXiv:221001690*, (2022).

22. Guo Q*, et al.* THz Time-Domain Spectroscopic Ellipsometry With Simultaneous Measurements of Orthogonal Polarizations. *IEEE Transactions on Terahertz Science and Technology* **9**, 422-429 (2019).

23. Chen X, Pickwell-MacPherson E. An introduction to terahertz time-domain spectroscopic ellipsometry. *APL Photonics* **7**, (2022).

24. Kamaraju N*, et al.* Subcycle control of terahertz waveform polarization using all-optically induced transient metamaterials. *Light: Science & Applications* **3**, e155-e155 (2014).

25. Choi WJ, Lee SH, Park BC, Kotov NA. Terahertz Circular Dichroism Spectroscopy of Molecular Assemblies and Nanostructures. *J Am Chem Soc* **144**, 22789-22804 (2022).

26. Baydin A*, et al.* Magnetic Control of Soft Chiral Phonons in PbTe. *Phys Rev Lett* **128**, 075901 (2022).

27. Cheng G, Choi WJ, Jang HJ, Kotov NA, Norris TB. Terahertz Time-Domain Polarimetry for Generalized Anisotropic and Chiral Materials. *Terahertz, Rf, Millimeter, and Submillimeter-Wave Technology and Applications Xii* **10917**, (2019).

28. Zhang ZY*, et al.* Terahertz circular dichroism sensing of living cancer cells based on microstructure sensor. *Anal Chim Acta* **1180**, (2021).

29. Leitenstorfer A, Hunsche S, Shah J, Nuss MC, Knox WH. Detectors and sources for ultrabroadband electro-optic sampling: Experiment and theory. *Appl Phys Lett* **74**, 1516-1518 (1999).

30. Wu Q, Zhang XC. Ultrafast electro-optic field sensors. *Appl Phys Lett* **68**, 1604-1606 (1996).

31. Wu Q, Zhang XC. 7 terahertz broadband GaP electro-optic sensor. *Appl Phys Lett* **70**, 1784-1786 (1997).

32. Kampfrath T, Notzold J, Wolf M. Sampling of broadband terahertz pulses with thick electro-optic crystals. *Appl Phys Lett* **90**, (2007).

33. Huber L, Maehrlein SF, Wang F, Liu Y, Zhu XY. The ultrafast Kerr effect in anisotropic and dispersive media. *J Chem Phys* **154**, 094202 (2021).

34. Zhang RX, Cui Y, Sun WF, Zhang Y. Polarization information for terahertz imaging. *Appl Optics* **47**, 6422-6427 (2008).

35. Nemoto N, Higuchi T, Kanda N, Konishi K, Kuwata-Gonokami M. Highly precise and accurate terahertz polarization measurements based on electro-optic sampling with polarization modulation of probe pulses. *Opt Express* **22**, 17915-17929 (2014).

36. van der Valk NCJ, van der Marel WAM, Planken PCM. Terahertz polarization imaging. *Optics Letters* **30**, 2802-2804 (2005).

37. Cornet M, Degert J, Abraham E, Freysz E. Terahertz Kerr effect in gallium phosphide crystal. *Journal of the Optical Society of America B* **31**, (2014).

38. Frenzel M, *et al.* Nonlinear terahertz control of the lead halide perovskite lattice. *Sci Adv* **9**, eadg3856 (2023).

39. Kaltenecker KJ, Kelleher EJR, Zhou B, Jepsen PU. Attenuation of THz Beams: A "How to" Tutorial. *Journal of Infrared, Millimeter, and Terahertz Waves* **40**, 878-904 (2019).

40. Zibod S, *et al.* Strong Nonlinear Response in Crystalline Quartz at THz Frequencies. *Advanced Optical Materials*, (2023).

41. Wei YX, Le JM, Huang L, Tian CS. Efficient generation of intense broadband terahertz pulses from quartz. *Appl Phys Lett* **122**, (2023).

42. Balos V, Wolf M, Kovalev S, Sajadi M. Optical rectification and electro-optic sampling in quartz. *Optics Express* **31**, (2023).

43. Rosner RD, Turner EH, Kaminow IP. Clamped Electrooptic Coefficients of Kdp and Quartz. *Appl Optics* **6**, 778-& (1967).

44. Shan J, Dadap JI, Heinz TF. Circularly polarized light in the single-cycle limit: the nature of highly polychromatic radiation of defined polarization. *Optics Express* **17**, 7431-7439 (2009).

45. Ghosh G. Dispersion-equation coefficients for the refractive index and birefringence of calcite and quartz crystals. *Opt Commun* **163**, 95-102 (1999).

46. Faust WL, Henry CH. Mixing of Visible and Near-Resonance Infrared Light in GaP. *Physical Review Letters* **17**, 1265-1268 (1966).

47. Eden DD, Thiess GH. Measurement of the Direct Electro-Optic Effect in Quartz at Uhf. *Appl Optics* **2**, 868-869 (1963).

48. Thämer M, Garling T, Campen RK, Wolf M. Quantitative determination of the nonlinear bulk and surface response from alpha-quartz using phase sensitive SFG spectroscopy. *The Journal of Chemical Physics* **151**, (2019).

49. Terki F, Levelut C, Boissier M, Pelous J. Low-frequency dynamics and medium-range order in vitreous silica. *Physical Review B* **53**, 2411-2418 (1996).

50. Buchenau U, Nücker N, Dianoux AJ. Neutron Scattering Study of the Low-Frequency Vibrations in Vitreous Silica. *Physical Review Letters* **53**, 2316-2319 (1984).

51. Naftaly M, Gregory A. Terahertz and Microwave Optical Properties of Single-Crystal Quartz and Vitreous Silica and the Behavior of the Boson Peak. *Applied Sciences* **11**, (2021).

52. Boyd RW. Nonlinear Optics, 3rd Edition. *Nonlinear Optics, 3rd Edition*, 1-613 (2008).

53. Tauchert SR, *et al.* Polarized phonons carry angular momentum in ultrafast demagnetization. *Nature* **602**, 73-77 (2022).

54. Castro-Camus E, Johnston MB. Extraction of the anisotropic dielectric properties of materials from polarization-resolved terahertz time-domain spectra. *Journal of Optics A: Pure and Applied Optics* **11**, (2009).

55. Cummings KD, Tanner DB. Far-Infrared Ordinary-Ray Optical-Constants of Quartz. *J Opt Soc Am* **70**, 123-126 (1980).

56. Choi WJ, *et al.* Chiral phonons in microcrystals and nanofibrils of biomolecules. *Nature Photonics* **16**, 366-373 (2022).

57. Davies CS, Fennema FGN, Tsukamoto A, Razdolski I, Kimel AV, Kirilyuk A. Phononic Switching of Magnetization by the Ultrafast Barnett Effect. *arXiv:230511551*, (2023).

58. Chen M, *et al.* Terahertz-slicing - an all-optical synchronization for 4(th) generation light sources. *Opt Express* **30**, 26955-26966 (2022).

# Supplementary Information

# for

# "Quartz as an Accurate High-Field Low-Cost THz Helicity Detector"

Maximilian Frenzel[1], Joanna M. Urban[1], Leona Nest[1], Tobias Kampfrath[1,2],
Michael S. Spencer[1], Sebastian F. Maehrlein[1,†]

[1] *Fritz Haber Institute of the Max Planck Society,*
   *Department of Physical Chemistry, 14195 Berlin, Germany*

[2] *Freie Universität Berlin, Department of Physics, 14195 Berlin, Germany*

†Corresponding author. Email: maehrlein@fhi-berlin.mpg.de

**Supplementary Note 1: Estimation of the THz peak field**

The THz pulse energy $E_p$ is calculated from measured THz power $P$ and laser repetition rate $f_{rep} = 1$ kHz by $E_p = P/f_{rep}$. The peak THz intensity $I_{peak}$ is determined from the THz pulse energy $E_p$, THz pulse duration (FWHM of THz intensity) $\Delta t$, and THz focus diameter $w$ (FWHM of THz intensity) using:

$$I_{peak} = \frac{E_p}{\Delta t} \cdot \frac{2\sqrt{\log 2}}{\sqrt{\pi}} \cdot \frac{2}{\pi \left(\frac{w}{2}\right)^2}. \qquad (\text{ S1 })$$

This expression assumes a Gaussian shape for both the spatial focus and temporal pulse form. Here, $w$ is estimated to be 350 µm using a FLIR A35 THz camera, and THz power $P$ was measured to be 1.65 mW with an Ophir Vega power meter with a 3A-P-THz sensor. The THz pulse duration is estimated from the FWHM of the THz intensity envelope $I_{env}(t)$ as determined using an EOS trace measured with quartz (see Fig. 3a).

The THz peak field strength $E_{\text{peak}}$ can finally be estimated using $E_{\text{peak}} = \sqrt{\frac{2}{\epsilon_0 c} I_{\text{peak}}}$. The estimated $E_{\text{peak}}$ in Fig. 3a used for calculating $r_{11}$ of z-cut quartz is therefore 1.04 MV/cm.

## Supplementary Note 2: Relation between THz-induced birefringence and electrooptic tensor elements of quartz

Electrooptic sampling is often treated from the viewpoint of a pump-induced change of birefringence that is experienced by a probe (or sampling) pulse in the material. For this purpose, it is useful to consider the material's index ellipsoid to describe its optical properties, which can be expressed in the general form:

$$\left(\frac{1}{n^2}\right)_1 x^2 + \left(\frac{1}{n^2}\right)_2 y^2 + \left(\frac{1}{n^2}\right)_3 z^2 + 2\left(\frac{1}{n^2}\right)_4 yz + 2\left(\frac{1}{n^2}\right)_5 xz + 2\left(\frac{1}{n^2}\right)_6 xy = 1 \qquad (\text{ S2 })$$

For z-cut α-quartz, belonging to the $D_3$ point group and 32 symmetry class[1,2], and oriented along the principle axes, the induced change to the index ellipsoid by a THz electric field $E_{\text{THz}}$ may then be written as:

$$\begin{bmatrix} \Delta(1/n^2)_1 \\ \Delta(1/n^2)_2 \\ \Delta(1/n^2)_3 \\ \Delta(1/n^2)_4 \\ \Delta(1/n^2)_5 \\ \Delta(1/n^2)_6 \end{bmatrix} = \begin{bmatrix} r_{11} & 0 & 0 \\ -r_{11} & 0 & 0 \\ 0 & 0 & 0 \\ r_{41} & 0 & 0 \\ 0 & -r_{41} & 0 \\ 0 & -r_{11} & 0 \end{bmatrix} \begin{bmatrix} E_{\text{THz},1} \\ E_{\text{THz},2} \\ E_{\text{THz},3} \end{bmatrix}, \qquad (\text{ S3 })$$

where $r_{ij}$ is the electrooptic tensor. For our experimental configuration, the sampling pulse field's and THz pulse field's polarizations are in the x-y plane and the z components of these fields are zero. We now consider the case, where the THz field is polarized along the y-axis: $E_{\text{THz},2} = t_{12}E_{\text{THz}}$; $E_{\text{THz},1} = E_{\text{THz},3} = 0$. Here, $E_{\text{THz},2}$ is already the field inside the sample and related to the incident THz field $E_{\text{THz}}$ via the Fresnel transmission coefficient, $t_{12} = (2/(1 + n_{\text{THz}}))$, if we neglect multiple internal reflections and absorption.

Note that the nonlinear susceptibility tensor $d_{ij}$ (methods section of main text) and $r_{ij}$ are related via $r_{ij} = -4d_{ji}/n^4$. Since $d_{11} \gg d_{14}$, we can assume $r_{11} \gg r_{41}$, and the resulting index ellipsoid becomes:

$$\frac{x^2}{n_o^2} + \frac{y^2}{n_o^2} + \frac{z^2}{n_e^2} - 2r_{11}t_{12}E_{\text{THz}}xy = 1, \qquad (\text{ S4 })$$

where $n_o$ and $n_e$ are the ordinary and extraordinary refractive indices of quartz respectively. We change to the coordinate system $(x', y', z)$, where $\hat{x}' = (\hat{x} + \hat{y})/\sqrt{2}$ and $\hat{y}' = (\hat{y} - \hat{x})/\sqrt{2}$. In this new coordinate system, the index ellipsoid becomes:

$$\frac{x'^2}{n_o^2}(1 + n_o^2 r_{11} t_{12} E_{THz}) + \frac{y'^2}{n_o^2}(1 - n_o^2 r_{11} t_{12} E_{THz}) + \frac{z^2}{n_e^2} = 1. \tag{S5}$$

The change of refractive index $\Delta n$ along x'- and y'-axis induced by a THz pulse polarized along the y-axis is therefore:

$$\Delta n_{x'} \approx -\frac{1}{2} n_o^3 r_{11} t_{12} E_{THz} \tag{S6}$$

$$\Delta n_{y'} \approx \frac{1}{2} n_o^3 r_{11} t_{12} E_{THz} \tag{S7}$$

A sampling pulse with frequency $\omega_s$ is polarized along the y-axis before it enters the EOS crystal with thickness $d$. The polarization state of the sampling pulse can thus be expressed as $\mathbf{E}_s = E_s \hat{\boldsymbol{y}} = \frac{E_s}{\sqrt{2}}(\hat{\boldsymbol{x}}' + \hat{\boldsymbol{y}}')$.

After passing through the EOS crystal and experiencing $\Delta n_{x'}$ and $\Delta n_{y'}$, the sampling pulse polarization is:

$$\mathbf{E}_s = \frac{E_s}{\sqrt{2}}\big(\exp(i\Delta\phi)\,\hat{\boldsymbol{x}}' + \exp(-i\Delta\phi)\,\hat{\boldsymbol{y}}'\big)$$
$$= \frac{E_s}{2}\big((\exp(i\Delta\phi) - \exp(-i\Delta\phi))\,\hat{\boldsymbol{x}} + (\exp(i\Delta\phi) + \exp(-i\Delta\phi))\,\hat{\boldsymbol{y}}\big), \tag{S8}$$

where $\Delta\phi = \frac{1}{2} n_o^3 r_{11} t_{12} E_{THz}\omega_s d/c$ under the assumption of zero phase-mismatch between sampling- and THz pulse and no absorption. The sampling pulse then passes through a $\lambda/4$ plate, so that $\mathbf{E}_s = iE_s\exp(-\frac{i\pi}{4})(\sin(\Delta\phi)\hat{\boldsymbol{x}} + \cos(\Delta\phi)\hat{\boldsymbol{y}})$. The following $\lambda/2$ plate and Wollaston prism spatially separate perpendicular polarization components in the $(x', y', z)$ basis, so that the measured photodiode intensities can be decomposed by $I_1 \propto |\hat{\boldsymbol{x}}' \cdot \mathbf{E}_s|^2$ and $I_2 \propto |\hat{\boldsymbol{y}}' \cdot \mathbf{E}_s|^2$.

The measured EOS signal in z-cut quartz is therefore:

$$S = \frac{I_1 - I_2}{I_1 + I_2} = \frac{E_s^2}{2E_s^2}((\sin(\Delta\phi) + \cos(\Delta\phi))^2 - (-\sin(\Delta\phi) + \cos(\Delta\phi))^2) \approx 2\Delta\phi$$
$$= \frac{n_o^3 r_{11} t_{12} E_{THz}\omega_s d}{c}, \tag{S9}$$

where we assumed a small $\Delta\phi \ll 1$ and used the small angle approximation ($\sin(\Delta\phi) \approx \Delta\phi$ and $\cos(\Delta\phi) \approx 1$). Note that the same expression for $S$ can be derived using the nonlinear polarization $\mathbf{P}^{(2)}$ description in the methods section of the main text.

**Supplementary Note 3: Comparison between quartz, ZnTe, and GaP EOS detectors**

Table S1 shows the key parameters that are relevant to assess the sensitivity and bandwidth of quartz, ZnTe, and GaP detectors. Note that the electrooptic coefficient $r_{eff}$ is not the best metric to describe the sensitivity of an EOS THz detector, since it does not include the effect of THz reflection, which means that less THz field is available for signal generation in the crystal. In addition, the measured EOS signal is also dependent on the third power of the refractive index $n_s^3$ experienced by the sampling pulse (see Eq. (S9)). A better metric for the THz detector sensitivity is therefore the figure of merit (FoM) defined as:

$$\text{FoM} = t_{12} r_{eff} n_s^3 = 2 r_{eff} n_s^3 / (1 + n_{THz}).$$  ( S10 )

| Detector | $r_{eff}$ (pm/V) | $n_{THz}$ | $n_s$ | GVM (ps/mm) | FoM (pm/V) |
|----------|------------------|-----------|-------|-------------|------------|
| ZnTe (110) | $r_{41} = 4$ (Ref. 3) | 3.18 (Ref. 3) | 2.85 (Ref. 3) | 1.1 (Ref. 3) | ~45 |
| GaP (110) | $r_{41} = 1.6$ (Ref. 4) | 3.70 (Ref. 4) | 3.35 (Ref. 4) | 1.2 | ~26 |
| Quartz (001) | $r_{11} = 0.1$ (this work) | 2.09 (Ref. 5) | 1.54 (Ref. 6) | 1.8 | ~0.2 |

**Table S1 | Overview of performance-related parameters for quartz, ZnTe, and GaP detectors**

**Supplementary Figures:**

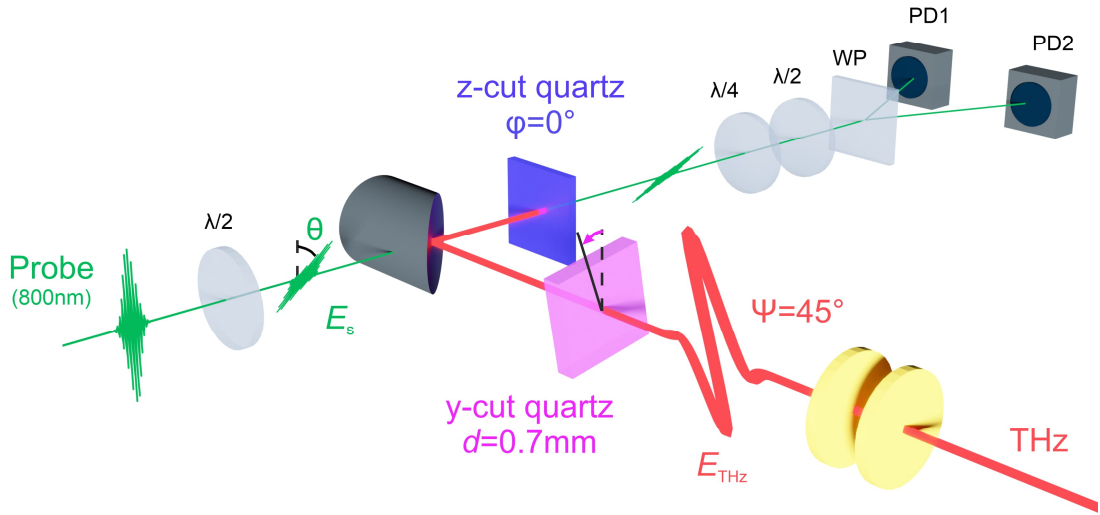

**Fig. S1 | Experimental setup for THz transmission of y-cut quartz.** THz electric field, which is initially polarized at $\psi = 45°$, passes through 0.7 mm crystalline y-cut quartz and is detected via 2D-EOS in 50 μm z-cut quartz for different azimuthal orientations of the y-cut quartz crystal (indicated by pink arrow).

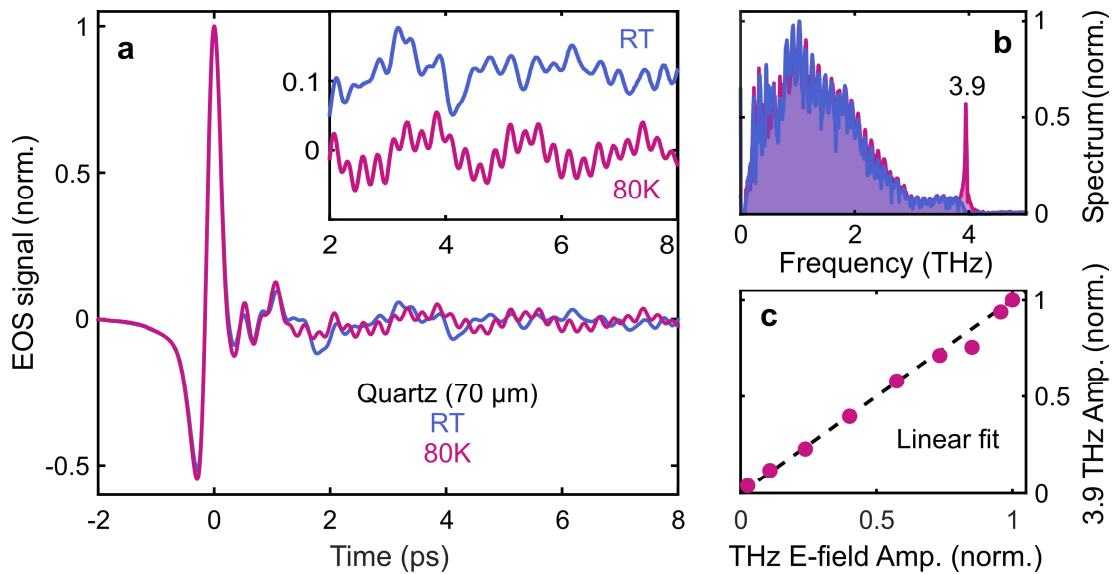

**Fig. S2 | EOS in quartz at room temperature and 80K. a** Time-domain EOS in quartz (70 μm thickness) measured at room temperature and 80 K. **b** Fourier transform of the respective signals, demonstrating that the EOS response at 80 K is similar to room temperature apart from the enhanced 3.9 THz TO phonon contribution at low temperature. Note that the frequency modulations are due to the Fabry-Pérot resonances from the windows of the optical cryostat used for these measurements. **c** 3.9 THz phonon amplitude scales linearly with THz electric field strength and can therefore be captured by linear response theory.

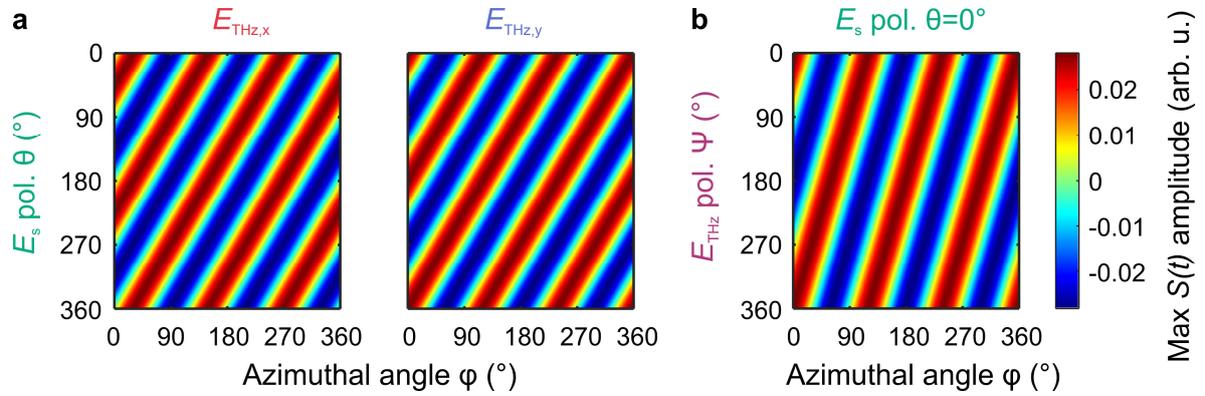

**Fig. S3 | Calculated quartz EOS dependence on azimuthal angle $\phi$, sampling pulse polarization $\theta$ and linear THz polarization $\psi$. a** Calculation for all possible $\phi$ and $\theta$ angles for fixed THz electric field x- and y- components, $\psi = 90°$ and $\psi = 0°$, respectively. It is evident that the EOS signal obeys a 3-fold symmetry in $\phi$, and a 2-fold symmetry in $\theta$. **b** Calculated EOS signal for all possible $\phi$ and $\psi$ angles at the fixed sampling pulse polarization $\theta = 0°$.

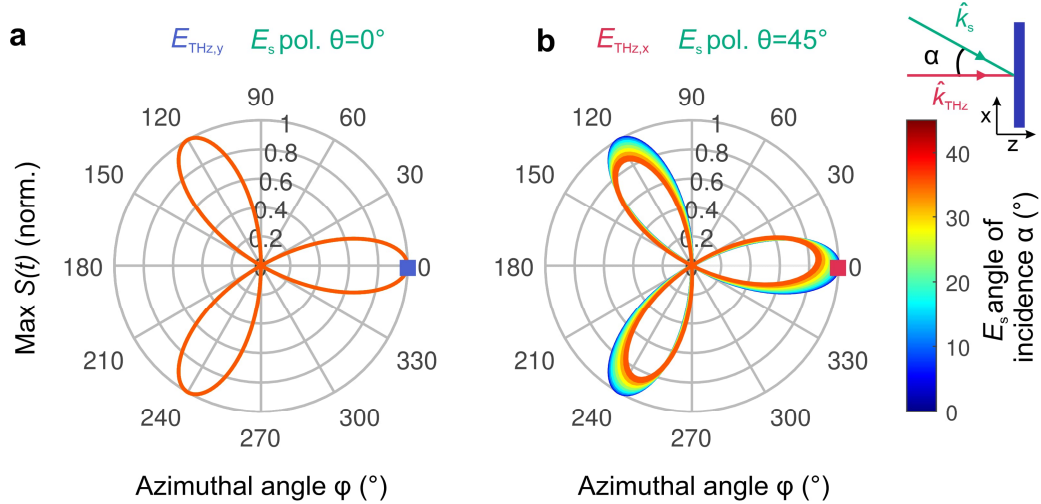

**Fig. S4 | Calculated quartz EOS azimuthal $\phi$ dependence as a function of sampling pulse angle of incidence $\alpha$ for the two geometries relevant for 2D-EOS. a** Calculation for all possible $\phi$ angles for fixed THz electric field y-component and sampling pulse polarization $\theta = 0°$ as a function of sampling pulse angle of incidence $\alpha$ between 0° and 45°. This plot illustrates that the THz y-component can be extracted (blue square) even at large angles of incidence with proper adjustment of the quartz response function. **b** Corresponding calculation for fixed THz electric field x-component and sampling pulse polarization at $\theta = 45°$, highlighting that the THz x-component may also still be extracted (red square) with the proper calibration of the EOS sensitivity.

**References:**

1.    Boyd RW. Nonlinear Optics, 3rd Edition. *Nonlinear Optics, 3rd Edition*, 1-613 (2008).
2.    Sutherland RL, McLean DG, Kirkpatrick S. *Handbook of nonlinear optics*, 2nd edn. Marcel Dekker (2003).
3.    Wu Q, Zhang XC. Ultrafast electro-optic field sensors. *Appl Phys Lett* **68**, 1604-1606 (1996).
4.    Wu Q, Zhang XC. 7 terahertz broadband GaP electro-optic sensor. *Appl Phys Lett* **70**, 1784-1786 (1997).
5.    Davies CL, Patel JB, Xia CQ, Herz LM, Johnston MB. Temperature-Dependent Refractive Index of Quartz at Terahertz Frequencies. *Journal of Infrared, Millimeter, and Terahertz Waves* **39**, 1236-1248 (2018).
6.    Ghosh G. Dispersion-equation coefficients for the refractive index and birefringence of calcite and quartz crystals. *Opt Commun* **163**, 95-102 (1999).